\documentclass[12pt]{article}
\usepackage{epsfig}
\usepackage{amsfonts}
\usepackage{amsopn}
\usepackage{amsmath}

\title{
\vskip -50 pt
\begin{flushright}
\normalsize\rm NORDITA-2011-23
\end{flushright}
\vskip 20 pt
Dirac membrane and hadrons}

\author{
Maciej Trzetrzelewski \thanks{e-mail: maciej.trzetrzelewski@gmail.com} \\ \\
NORDITA,  \\
Roslagstullsbacken 23, 106 91 Stockholm, \\
Sweden}

\begin{document}
\date{}
\maketitle

\abstract{In $c=1$ units the product (mass
$\times$ radius) for the neutron and the proton is about $4.7\hbar$ assuming their radii equal to 1fm. We show that the corresponding products for
the Dirac neutral and charged membrane coincide and are equal $1.6\hbar$.}

\section{Introduction}

In 1962 Dirac, trying to explain the value of the muon mass,
considered an idea that the electron could be modeled by a
conducting, elastic membrane of spherical topology \cite{Dirac} (at
the same time he also considered a neutral membrane in the presence
of the gravitational field \cite{jablonna} - the finite size of the
electron was originally considered much earlier by Lorentz, Abraham,
Bucherer and Langevin, see \cite{Wroblewski} and references therein). Assuming the Lagrangian
$L=L_{EM}+L_{mem.}$, where $L_{EM}$ is a usual term for the
electromagnetic field and $L_{mem.}$ is given by the membrane
world-volume, it can be shown that for the spherically symmetric
case with the radial coordinate $\rho$, the Hamiltonian of the model
(in $c=1$ units) is given by
\begin{equation} \label{1}
H_r=\sqrt{-\hbar^2\partial_{\rho}^2+\omega^2 \rho^4 } +\frac{e^2}{2\rho}
\end{equation}
where $\omega/4\pi$ is the membrane tension and $e$ is the electric
charge. Using the Bohr-Sommerfeld quantization method, one finds
that the first excitation of the membrane corresponds to the energy
$\approx 53m_e$, where $m_e$ is the mass of the electron. Dirac
notes that in order to get a closer value to the experimental
$m_{\mu}\approx 207m_e$ one would presumably have to introduce spin
into the theory by performing the square root appearing in $H_r$.
While this might be true, in this paper we focus on another possible
application of Dirac's membrane i.e. the effective description of
hadrons. We find a surprisingly good agreement between the product
(mass$\times$radius) for charged/netral membrane and the
experimental values for the proton/neutron respectively.

\section{Dirac's model}

Among all closed (compact, without the boundary) objects coupled to
the electromagnetic field $A_{\mu}$ in $D=4$ spacetime dimensions,
membranes play a special role ensuring that $A_{\mu}$ is nowhere
singular. The reason for this lying in the fact that such object
split $\mathbb{R}^3$ into two disconnected regions: the interior and
the exterior. Equations of motion for membranes are nonlinear hence quite
complicated and unlike in the case of points or strings very little
is known about the exact solutions (for some examples see
\cite{solutions}). As for the quantum theory, the existing method of
quantizing the bosonic membrane \cite{hoppephd} is highly
non-trivial resulting in a certain quantum mechanical model with
matrix degrees of freedom. When the size of a matrix is taken to
infinity the precise description of the quantum membrane is
obtained. In spite of these complications recently there has been a
considerable progress in understanding the bosonic membrane theory
\cite{recent}.

Since closed membranes in $\mathbb{R}^3$ provide a natural split of
space into the interior and the exterior there exists a proffered
curvilinear system $x^{\mu}$ in spacetime and a function $f(x)$ such
that the equation $f(x)=0$ describes a membrane and  equations
$f(x)>0$, $f(x)<0$ describe a region outside or inside the membrane,
respectively. It is convenient to fix the curvilinear system such
that $x^1=f(x)$ and choose $\sigma^0=x^0=:\tau$, $\sigma^1=x^2$,
$\sigma^2=x^3$ for the three variables $\sigma^{\alpha}$,
$\alpha=0,1,2$ - the internal parametrization of the membrane
word-volume. The action for a membrane coupled to the
electromagnetic field considered by Dirac \cite{Dirac} is (in $c=1$
units)
\begin{equation} \label{diraction}
S=S_{EM} +S_{mem.},
\end{equation}
\[
S_{EM}=-\frac{1}{16\pi}\int_{x^1>0} J g^{\mu\rho}g^{\nu\sigma}F_{\mu\nu}F^{\rho\sigma}d^4x, \ \ \ F_{\mu\nu}=\partial_{[\mu}A_{\mu]},
\]
\[
S_{mem.}= - \frac{\omega}{4\pi}\int_{x^1=0} M dx^0dx^2dx^3
\]
where $g^{\mu\nu}$ is the metric corresponding to the curvilinear
system $x^{\mu}$ (concretely Dirac takes the induced metric
$g_{\mu\nu}=\partial_{\mu}y^{\Lambda}\partial_{\nu}y_{\Lambda}$,
$\Lambda=0,1,2,3$, where $y^{\Lambda}$ are rectilinear and
orthogonal), $J=\sqrt{-\det g_{\mu\nu}}$ and $M=J\sqrt{-g^{11}}$.
The coupling is due to the factor $J$ appearing in both $S_{EM}$ and
$S_{mem.}$. Varying (\ref{diraction}) with respect to
$y^{\Lambda}$ one arrives at the equations of motion which for the
spherically symmetric case $x^1=r-\rho$, $x^2=\theta$, $x^3=\phi$
(so that the surface is given by $x^1=0$) give
\begin{equation} \label{eomr}
\frac{d}{dt} \frac{\dot{\rho}}{\sqrt{1-\dot{\rho}^2}}=\frac{e^2}{2\omega \rho^4} - \frac{2}{\rho\sqrt{1-\dot{\rho}^2}}.
\end{equation}
The balance between the repulsive electromagnetic forces and the
attractive ones, due to the positive membrane tension, is when
$\dot{\rho}=0$ hence $a^3=e^2/4\omega$ where $a$ is the radius of
the electron. On the other hand the total energy of a system at rest
$E=e^2/2\rho + \beta \rho^2$ (minimal in the equilibrium provided
$\beta=\omega$) is equal to both $3e^2/4a$ and $m_e$ when $\rho=a$.
Therefore one concludes that $a=3e^2/4m_e=0.75r_e$ where $r_e$ is
the classical electron radius, $r_e \approx 2.8$fm (the value
$a=2.1$fm is of course not realistic as it is bigger then the charge
radius of the proton $0.87$fm \cite{pdg} - according to Dehmelt the
radius of the electron could be of order $10^{-7}$fm \cite{Dehmelt}).
Considering small oscillation about the equilibrium one finds the
corresponding frequency to be $\sqrt{6}/a$ hence the energy of one
quantum would be $h\nu = \sqrt{6}\hbar/a = 448m_e$.

To improve the analysis one proceeds to the hamiltonian formalism.
While the details of this are quite involved
the final answer for the spherically symmetric case
turns out to be particularly simple given by (\ref{1}). The complications are due to the choice of
the coordinates $x^{\mu}$ hence the loss of the explicit
diffeomorphism invariance of the action. There exists a generally
covariant formulation \cite{Barut} in which one considers, in addition to
$S_{EM}$ and $S_{mem.}$ written for an arbitrary $x^{\mu}$ and
$\sigma^{\alpha}$, an extra term proportional to
$e^{\alpha}\partial_{\alpha}X^{\mu}A_{\mu}$  where  $e^{\alpha}$ is
the current-charge density on the membrane. One then shows that all
Dirac's findings, in particular the hamiltonian (\ref{1}), can be
obtained in a less elaborate way keeping the explicit diffeomorphism
invariance.

Using the
Bohr-Sommerfeld quantization condition one
obtains an approximation which can be written as
\[
\frac{m_{Dirac}}{m_e}=\frac{1}{3}\left(\frac{32\sqrt{\pi} \Gamma(7/4)}{ \alpha
\Gamma(1/4)}n\right)^{\frac{2}{3}}
\]
where $\alpha=e^2/\hbar c$ is the fine structure constant, obtaining
\[
\frac{m_{Dirac}}{m_e}\approx 52.4, \ \ 83.1, \ \ 109.0, \ \ 132.0,  \ \  \ldots
\]
for consecutive values of $n$.

This approximation can be improved as observed in \cite{gnadig} by
noting that the in the $\alpha \to 0$ limit the coulomb term in
(\ref{1}) acts like an infinite wall and hence one should apply
different boundary conditions in the Bohr-Sommerfeld quantization
procedure. As a result one obtains
\[
\frac{m_{improved}}{m_e}\approx 43.3, \ \ 76.1, \ \ 102.9, \ \ 126.1, \ \  \ldots
\]
which agree with the numerical values also found in \cite{gnadig}.

\subsection{Further investigation}

Let us rewrite the hamiltonian $H_r$ in (\ref{1}) in dimensionless
variable $x=\rho/a$
\begin{equation} \label{hr}
H_r= \frac{4m_e}{3\alpha}\left(\sqrt{-\partial_x^2+\frac{\alpha^2}{16}x^4} +\frac{\alpha}{2x} \right)
\end{equation}
where we used $\omega=\frac{e^2}{4a^3}$, $a=\frac{3e^2}{4m_e}$.

While finding an analytic expression for the square root of the
positive differential operator $H$ is in general difficult the calculation
of the matrix elements of the square root $\sqrt{H}$ is possible
using the standard procedure by diagonalizing $H$ and taking
$\sqrt{H}=U^T\sqrt{E}U$ where $U$ is s.t. $H=U^TEU$,
$E=diag(E_n)$. Applying numerical techniques (for details see the
Appendix  - we use a different technique compared to \cite{gnadig})
we find that this prescription gives
\[
\frac{m_{our}}{m_e}\approx 43.6, \ \ 76.3, \ \ 103.0, \ \ 126.5, \ \  \ldots
\]
which are consistent with the results of \cite{gnadig}.

In order to find the radius of the first two excitations (let us
denote the corresponding wavefunctions by $\psi_{\mu}$ for
$E_1=43.6m_e$ and $\psi_{\tau}$ for $E_2=76.3m_e$) we use their probability densities (Figure 1).
\begin{figure}[h]
\centering
\leavevmode
\epsfverbosetrue
\epsffile{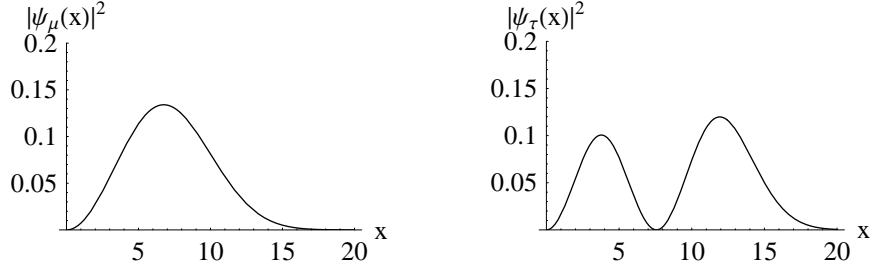}
\caption{ Probability densities $|\psi_{\mu}(x)|^2$ and $|\psi_{\tau}(x)|^2$.}
\end{figure}
It follows that $|\psi_{\mu}(\rho)|^2$ has a maximum for $\rho\approx
6.7a$ while $|\psi_{\tau}(\rho)|^2$ has two maxima for $\rho\approx
3.8a$ and $\rho\approx 11.9a$ so that the probability density $|\psi_{\tau}(\rho)|^2$ will be viewed in $\mathbb{R}^3$ as two concentric membranes.

It is somewhat interesting to consider the generalization of Dirac
action to the case of two (or more) concentric membranes. We will
assume that the total charge $e$ is on the exterior membrane. If the
tension of the inner, neutral shell is $k \omega$, $k>0$ the
corresponding hamiltonian of this system would be
\begin{equation} \label{1a}
H_{r\rho}=\sqrt{-\partial_r^2+k^2\omega^2 r^4 } + \sqrt{-\partial_{\rho}^2+\omega^2 \rho^4 } +\frac{e^2}{2\rho}.
\end{equation}
The eigenvalues of this hamiltonian are lifted compared to $H_r$.
One can choose $k$ so that its first excitation is equal to the mass
of the muon (this takes place for $k \approx 63$) however the second
excitation is not even close to the mass of the taon. Of course one
could consider three concentric membranes and fit the additional
tension so that the mass of the taon appears but such model would be
just a fit.

\section{Reinterpretation}
Perhaps the result $m_{\mu}=43.6m_e$ could be improved by
introducing spin in some clever way into the theory. However it seems that this membrane model works surprisingly well if we identity
charged membrane with the proton and the neutral membrane with the
neutron.

Let us start with the case of the neutral membrane. The discreteness
of the hamiltonian (\ref{1}) is not due to the Coulomb term.
Classically, without the electromagnetic field the membrane will
collapse (i.e. $\rho=0$ after a finite time - $\rho(\tau)$ can be
obtained from the equations of motion which imply
$\ddot{\rho}\rho=2(\dot{\rho}^2-1)$ solved by the Jacobi sine
function) but the operator
\begin{equation} \label{hneutral}
H_{neut.}=\sqrt{-\hbar^2\partial_{\rho}^2+\Omega^2 \rho^4 }=(\hbar^2 \Omega)^{1/3}\sqrt{-\partial_y^2+ y^4 }
\end{equation}
where $y=\rho/(\hbar/\Omega)^{1/3}$ is dimensionless, itself is discrete (since it is a square root of a discrete,
positive definite operator) hence quantum mechanically a free
membrane will develop bound sates. The natural unit of energy is now
$(\hbar^2 \Omega)^{1/3}$ where the tension $\Omega$ is not specified. The
hamiltonian $H_{neut.}$ has parity even and odd eigenstates. The parity even states
have the maximum of the probability density for $\rho=0$ which corresponds to a membrane with $0$ radius - a point.
However the parity odd states have the maximum for $\rho>0$ for which the eigenvalues are
\[
\frac{m_{neut.}}{(\hbar^2 \Omega)^{1/3}}\approx 1.95, \ \ 3.41, \ \ 4.61, \ \ 5.66 \ \ldots \ .
\]
The probability density of the first excitation $m_0\approx1.95(\hbar^2 \Omega)^{1/3}$ has a maximum at $\rho_0 \approx 0.82(\hbar/ \Omega)^{1/3}$ (Figure 2)
\begin{figure}[h]
\centering
\leavevmode
\epsfverbosetrue
\epsffile{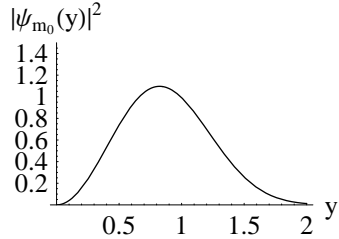}
\caption{First excitation for a neutral membrane.}
\end{figure}
hence $m_0 \rho_0 \approx 1.6\hbar$.

It is quite amusing that for the case of the neutron we have $m_{n}
\rho_{n} \approx 4.7 \hbar$ (in $c=1$ units, we took $\rho_{n} =1$fm) in rough
agreement with the result for the neutral membrane. Considering the
fact that we do not mention quarks, gluons and the spin, this naive picture of
a neutron as a neutral membrane works surprisingly well.

We can now do the analogous analysis for the charged membrane. The
hamiltonian is the same as for the case of the electron only written
in different units (cp. (\ref{hr}))
\[
H_{charged}=(4\hbar^2 \Omega/\alpha)^{1/3}\left(\sqrt{-\partial_z^2+ \frac{\alpha^2}{16}z^4 }+\frac{\alpha}{2z} \right)
\]
(where, $z=\rho/(\alpha\hbar/4\Omega)^{1/3}$, $\Omega$ can be
related to the classical radius at the equilibrium). Due to the
Coulomb term we must assume $\psi(0)=0$ for the wavefunctions hence
all the excitations have nonzero radius. The eigenvalues are
\[
\frac{m_{charged}}{(4\hbar^2 \Omega/\alpha)^{1/3}}\approx 0.239, \ \ 0.418,  \ \ 0.564, \ \ 0.693, \ \ldots \
\]
and the lowest energy state $m_0\approx0.239(4\hbar^2
\Omega/\alpha)^{1/3}$ is peaked at $\rho_0\approx6.75(\alpha\hbar/
4\Omega)^{1/3}$ (same as for $\psi_{\mu}$, see Figure 1) implying $m_0 \rho_0
\approx 1.6 \hbar$.

\section{Conclusions}
It is not a surprise that effective description of hadrons in terms
of bag-like models will essentially give results consistent with
experimental data. However what is striking about the Dirac membrane
model is its huge simplicity compared to the actual processes taking
place - no quarks, gluons, spin - one wonders why this model works
at all? Moreover, the value of the only dimensional parameter of the
theory, the tension $\Omega$, was nowhere used as it cancels out in
the calculation.  It seems that, at least for the purposes of
calculating the product (mass $\times$ radius), the spin can be
neglected while the strong forces replaced by a $2+1$ dimensional
field theory - signaling a sort of holographic \cite{thooft}
behavior.

\vspace{12pt} \noindent{\bf Acknowledgment}  I thank  M. Ku\'zniak
and P. O. Mazur for discussions and the correspondence as well as the Swedish Research Council and KTH for support.

\section*{Appendix}
In order to obtain numerically the eigenvalues of
\[
H_x:=\frac{4}{3\alpha}\left(\sqrt{\partial_x^2+\frac{\alpha^2}{16}x^4} +\frac{\alpha}{2x} \right)
\]
we use the cutoff method \cite{cutoff} which consists of
calculating the matrix elements of $H_x$ in some orthonormal basis
(on $[0,\infty)$ in our case), truncate the infinite matrix and then
numerically diagonalize it. The spectra and the eigenvectors of the
truncated matrices converge very quickly to their exact
counterparts.

An important step in this method is the choice of the convenient
basis (a common choice is the basis in the Fock space - not
suitable here). Due to the $1/x$ part in $H_x$ the regularity of the
wavefunction $\psi(x)$ at the origin implies that $\psi(x)\sim x^l$
for $l\ge 1$. For this reason we choose the orthonormal basis
$e_n(x)$, $n=1,2,3,\ldots$ as
\[
e_n(x)=\frac{1}{\sqrt{n(n+1)}}x L^{(2)}_{n-1}(x)e^{-x}, \ \ \ \int_0^{\infty}e_n(x)e_m(x)dx=\delta_{nm}
\]
where $L^{(2)}_n(x)$ are generalized Laguerre polynomials. In this basis the explicit
representation of operators $K=-\frac{d^2}{dx^2}$, $W=x^4$ and
$V=1/x$ can be obtained with
\[
K_{nm}:=(e_n,Ke_m), \ \ W_{nm}:=(e_n,We_m), \ \ V_{nm}:=(e_n,Ve_m).
\]
At this point we introduce a cutoff $N_{max}$, consider a finite
matrix  $h^{(N_{max})}_{nm}=K_{nm}+W_{nm}$, $n,m\le N_{max}$ and
then numerically find its eigenvalues $E^{(N_{max})}_k$ and the
matrix $U^{(N_{max})}$ s.t.
\[
h^{(N_{max})}=U^{(N_{max}) \ T}E^{(N_{max})} U^{(N_{max}) }, \ \ \ \
E^{(N_{max})}=diag(E^{(N_{max})}_k).
\]
The square root of
$h^{(N_{max})}$ can now be obtained and the
matrix representation for the overall operator $H_x$ is
\[
H_x^{(N_{max})}=U^{(N_{max}) \ T}\sqrt{E}^{(N_{max})} U^{(N_{max}) } + V^{(N_{max})}.
\]
The spectrum of $H_x^{(N_{max})}$ quickly converges to the exact values (Figure 3).
\begin{figure}[h]
\centering
\leavevmode
\epsfverbosetrue
\epsffile{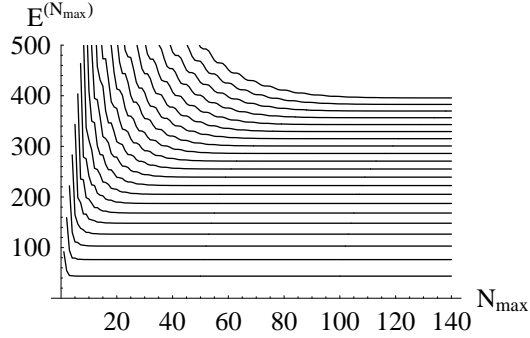}
\caption{ The convergence of the eigenvalues of $H_x^{(N_{max})}$.}
\end{figure}
To obtain the wavefunctions corresponding to $E^{(N_{max})}_n$ we
use the eigenvectors $v^{(N_{max})}_n$ of
$H_x^{(N_{max})}$ i.e.
\[
\psi^{(N_{max})}_n(x)=\sum_{k=1}^{N_{max}}[v^{(N_{max})}_n]_k e_k(x).
\]
The convergence of $\psi^{(N_{max})}_n(x)$ to the exact $\psi_n(x)$
is governed by the corresponding convergence of the eigenvectors. In
practice the cutoff$=10$ already gives very accurate approximation
for first excitations.

To find the eigenvalues of the parity even states of
(\ref{hneutral}) we use the same numerical method but with help of a
different, more convenient basis
\[
f_n(x)=\frac{1}{\sqrt{2^{n-1} n! \sqrt{\pi} )}} H_{2n-1}(x)e^{-x^2/2}, \ \ \ \int_0^{\infty}f_n(x)f_m(x)dx=\delta_{nm}
\]
where $H_n(x)$ are Hermite polynomials.

\end{document}